\documentclass[runningheads]{cl2emult}

\usepackage{makeidx}  
\usepackage{graphicx} 
\usepackage{subeqnar} 
\usepackage{multicol} 
\usepackage{cropmark} 
\usepackage{eso}      
\makeindex            
\usepackage{psfig}

\def\Msun{\,M$_\odot$}

\begin{document}
\title*{Formation of Black-Hole X-Ray Binaries with Low-Mass Donors}
\toctitle{Formation of Black-Hole X-Ray Binaries
\protect\newline with Low-Mass Donors}
%
%
\titlerunning{Black-Hole X-Ray Binaries}
%
\author{Vassiliki Kalogera}
\authorrunning{Vassiliki Kalogera}
%
%
\institute{Harvard-Smithsonian Center for Astrophysics, Cambridge MA
02138, USA}

\maketitle              

\begin{abstract}
 The characteristics of black-hole X-ray binaries can be used to obtain
information about their evolutionary history and the process of
black-hole formation. In this paper I focus on systems with donor masses
lower than the inferred black-hole masses.  Current models for the
evolution of hydrogen-rich, massive stars and of helium stars losing
mass in a wind cannot explain the current sample of black-hole mass
measurements. Assuming that the radial evolution of mass-losing massive
stars is at least qualitatively accurate, I show that the properties of
the BH companions lead to constraints on the masses of black-hole
progenitors (at most twice the black-hole mass) and on the strength of
winds in helium stars (fractional amount of mass lost smaller than about
50\%). Constraints on common-envelope evolution are also derived.  
\end{abstract}

\section{Introduction} 

Radial velocity measurements of the non-degenerate donors in X-ray
binaries combined with information about donor spectra and optical light
curves allow us to measure the masses of accreting compact
objects~\cite{C98}. At present, measured masses for nine X-ray
transients\index{X-ray transients} exceed the optimum maximum neutron
mass~\cite{KB96} and the binaries are thought to harbor black
holes\index{black holes} (BH). Studies of the properties of such systems
can shed a light to their evolutionary history and the process of BH
formation.

Black-hole X-ray transients correspond to low-mass X-ray binaries with
neutron stars (NS) in that mass transfer is driven by Roche-lobe
overflow and the donors are less massive than the BH. This critical
(maximum) mass ratio of about unity allows the donor to transfer mass
stably to the compact object. However, for the majority of the BH X-ray
transients (including six BH candidate systems based on their spectral
properties~\cite{C97}) the donors are less massive than $\sim 1$\Msun,
much less massive than the typical BH masses. Only two systems, J1655-40
and 4U1543-47, have donors of $\sim 2.3$\Msun\, and $\sim 2-3$\Msun,
respectively. This apparent paucity of intermediate-mass donors (more
massive than $\sim 2$\Msun) cannot be explained by mass-transfer
stability considerations alone.

In what follows, we address the issue of donor masses in BH binaries by
studying a larger set of constraints imposed on their progenitors. We
find that there is a strong dichotomy between the formation of systems
with low- and intermediate-mass donors. Formation of both at the
appropriate relative fraction requires little mass loss at BH formation,
weak helium-star winds\index{helium-star winds}, and possibly energy
sources other than the orbit for common-envelope\index{common-envelope
phase} (CE) ejection. This study and the results obtained are described
in more detail in~\cite{K99}.

\section{Evolutionary Constraints}

We consider BH X-ray binary formation from primordial binaries with
extreme mass ratios evolving through a CE phase, similar to
the formation channel for NS low-mass X-ray binaries, e.g.,~\cite{vdH83}.
The primary must be massive enough so that its helium core exposed at the
end of the CE phase collapses into a BH. The X-ray phase is
initiated when the donor fills its Roche lobe because of orbital shrinkage
through magnetic braking (for low-mass donors) or of radial expansion
through nuclear evolution on the main sequence (for intermediate-mass
donors).

Black-hole binary progenitors evolve through this path provided that the
following constraints are satisfied: 
 \begin{itemize}
 \item
 The orbit is small enough that the primary fills its Roche lobe and the
binary enters a CE phase. 
 \item
 At the end of the CE phase the orbit is wide enough so that both the
helium-rich primary and its companion fit within their Roche lobes. The
constraint for the companion turns out to be stricter.
 \item
 The system remains bound after the collapse of the helium star. In the
absence of kicks imparted to the BH, this sets an upper limit on the mass
of the BH progenitor. 
 \item
 After the collapse, the orbit must be small enough so that mass transfer
from the donor starts before it leaves the main sequence and within
10$^{10}$\,yr. 
 \item
 Mass transfer from the donor proceeds stably and at sub-Eddington
rates. This sets an upper limit to the donor mass on the zero-age main
sequence and to the orbital size for more evolved donors.
 \end{itemize}

\section{Donor Masses in Black-Hole X-ray Binaries}

For a specific value of the BH mass, the above constraints translate into
limits on the properties, circularized post-collapsed orbital sizes ($A$)  
and donor masses ($M_d$), of BH binaries with Roche-lobe filling
donors. The relative positions of these limits on the $A-M_d$ plane and
the resulting allowed $M_d$ ranges are exactly determined by three well
constrained model
parameters:
 \begin{itemize}
 \item 
 The amount of mass loss from the binary during BH formation,
characterized by the ratio $M_{He,f}/M_{\rm BH}$, where $M_{He,f}$ is the
mass of the helium-rich BH progenitor at the time of the collapse. For the
post-collapse system to remain bound it must be $1 \leq M_{He,f}/M_{\rm
BH} \leq 3$.
 \item
 The amount of mass lost in the helium-star wind between the end of the CE
phase and the BH formation, characterized by the ratio $M_{He,f}/M_{He}$,
where $M_{He}$ is the initial helium-star mass (at the end of the CE
phase). This ratio must lie in the range $0-1$. 
 \item
 The CE efficiency, $\alpha_{CE}$, defined as the ratio of the CE binding
energy to the orbital energy released during the spiral-in of the
companion. Although the absolute normalization of $\alpha_{CE}$ is not
well determined~\cite{K99}, values higher than unity imply the existence
of energy sources other than the orbit (ionization or nuclear burning
energy). 
 \end{itemize}

Note that the last two of the constraints depend {\em only} on the BH
mass, while $\alpha_{CE}$ affects {only} the upper limit on $A$ (first
of the constraints \S\,2). For different values of these three
parameters, the positions of the limits on the $A-M_d$ plane change and
three different outcomes with respect to the donor masses are possible:
BH binaries can be formed with (i) {\em only} low-mass; (ii) {\em only}
intermediate-mass; (iii) {\em both} low- and intermediate-mass donors.

 \begin{figure}
 \centering
 \includegraphics[width=0.5\textwidth,angle=-270.0]{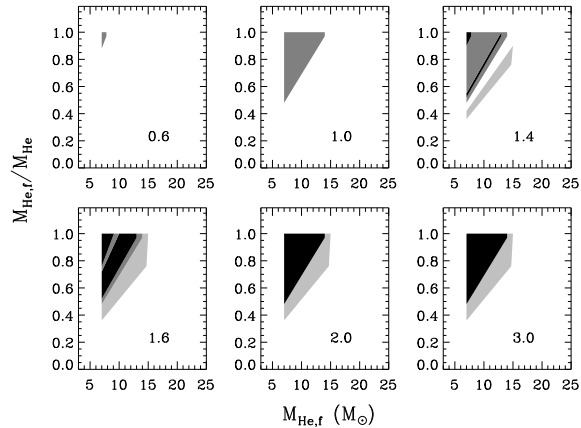}
 \caption[]{Limits on the parameter space of the final (pre-collapse)
helium-star mass, $M_{He,f}$, and the ratio, $M_{He,f}/M_{He}$, for six
values of the $\alpha_{\rm CE}=$0.6, 1.0, 1.4, 1.6, 2.0, 3.0, and for a
7\Msun\, BH. Conditions in the unshaded areas do not allow the formation
of BH binaries with main-sequence Roche-lobe filling donors; conditions
in the light-gray, dark-gray, and black areas allow the formation of
systems with only low-mass, only intermediate-mass, and both types of
donors, respectively.} 
 \end{figure}

The donor types as a function of the three parameters, $M_{He,f}$,
$M_{He,f}/M_{He}$, and $\alpha_{\rm CE}$, are shown in Fig.~1, for a
7\Msun\, BH. For $\alpha_{\rm CE}$ smaller than $\sim 0.5$, the orbital
contraction is so high that the donor stars cannot fit in the post-CE
orbits, and hence no BH X-ray binaries are formed. As $\alpha_{\rm CE}$
increases, CE ejection without the need of strong orbital contraction
becomes possible for the more massive of the donors, while formation of
binaries with low-mass donors occurs only if $\alpha_{\rm CE}>1.5$. The
results become independent of $\alpha_{\rm CE}$ for values in excess of
$\sim 2$, when the upper limit for CE evolution (first of the
constraints) lies at high enough values of $A$ that it never interferes
with the other limits.

The dependence of these results on the two mass-loss parameters (wind
anf collapse) are determined by their association with orbital
expansion. For strong helium-star wind mass loss
($M_{He,f}/M_{He}<0.35$), the progenitor orbits expand so much that
donors less massive than the BH can never fill their Roche lobes on the
main sequence. Both low- and intermediate-mass donors are formed only if
less than 50\% of the initial helium-star mass is lost in the wind. Mass
loss at BH formation is limited to BH progenitors less massive than
about twice the BH mass so that post-collapse systems with low-mass
donors remain bound.

\begin{figure}
 \centering
 \includegraphics[width=0.35\textwidth,angle=-270.0]{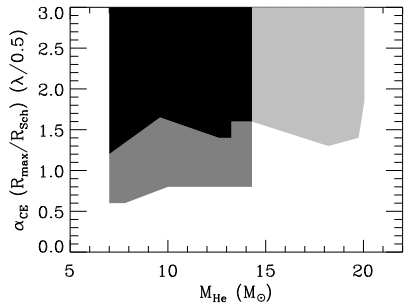}
 \caption[]{Limits on the parameter space of the initial (post-CE)
helium-star mass, $M_{He}$, and the common-envelope efficiency,
$\alpha_{\rm CE}$, properly normalized (by the maximum stellar radii of
massive stars~\cite{S92} and the central-concentration parameter,
$\lambda$), for a 7\Msun\, BH. Shade coding is as in Figure 1. }
 \end{figure}

The dependence on $M_{He,f}$ of the orbital expansion during helium-star
wind mass loss and BH formation is such that the ratio of circularized
post-collapse over post-CE orbital separations becomes independent of
$M_{He,f}$. This means that, for a specific BH mass, the position of the
limits on the $A-M_d$ plane depend only on the initial helium-star mass
and the CE efficiency. Indeed, in Fig.~1, the change of donor types
occurs along straight lines in the $M_{He,f}/M_{He}$--$M_{He,f}$ plane,
or else along lines of constant $M_{He}$. This simplifying property
allows us to combine the panels in Fig.~1 into one plot (Fig.~2). It is
evident that formation of 7\Msun\, BH X-ray binaries with both low- and
intermediate-mass donors (as required by the observed sample) constrains
the common-envelope efficiency to relatively high values and the initial
helium-star progenitors at most twice as massive as the BH
(corresponding to initial primaries in the range 25-45\Msun).

Additional constraints can be obtained by examining the relative numbers
of systems with low- and intermediate-mass donors formed for the
parameters in the black-shaded areas in Figs.\ 1 and 2.  The lifetimes
for the two different types are determined by the process that drives
mass transfer. The magnetic-braking timescale, for low-mass donors is
comparable to the nuclear evolution timescale of intermediate-mass
stars~\cite{K99}. The number ratio then becomes equal to the ratio of
birth rates. The latter can be calculated using the derived limits on
$A$ and $M_d$ and assumed distributions of mass ratios and orbital
separations of primordial binaries. The results indicate that even when
low-mass companions in primordial binaries are strongly favored, BH
binaries with intermediate-mass donors are much more easily formed
because of the larger range of orbital separations allowed to their
progenitors. Models predict a paucity of intermediate-mass donors only
for rather high $\alpha_{\rm CE}$ values ($>3$) or for moderate
$\alpha_{\rm CE}$ values ($1.5-2$) and BH progenitors either slightly
more massive or twice as massive as the BH.

\section{Discussion}

We find that the models for BH formation are consistent with the
properties of the observed sample if (i) wind mass-loss from helium
stars is limited so that they lose less than half of their initial mass
(ii) helium stars that form black holes are at most twice 
more massive than the black holes, and (iii) CE efficiencies are
relatively high and, depending on the exact radii of massive stars and
their density profiles, significant contributions from energy sources
other than the orbit may be required. Note that amounts of mass lost in
helium-star winds and in BH formation are actually anti-correlated. If
one is close to the maximum allowed then the other must be minimal (see
Fig.\ 1). These results are quite robust and do not depend on the
assumed BH mass nor the properties of primordial binaries. 

The present study allows us to place constraints on the extent of wind
mass loss from helium stars as they evolve towards collapse, primarily
because helium cores are exposed at the end of the CE phase {\em prior}
to core helium exhaustion. Current models of helium-star evolution
through core helium burning~\cite{W95} predict amounts of mass lost in
the wind significantly larger than the maxima allowed for BH X-ray
transient formation ($<50$\%). In fact, the final helium-star masses in
these models are $\sim 4$\Msun, far too small to explain the BH mass
measurements. Therefore, if the CE phase is initiated early in the core
helium burning phase of the primary, then helium-star winds must be much
weaker than thought until now. It is noteworthy that more recent
empirical estimates of wind mass loss rates~\cite{HK98} show a downward
trend.

The strength of helium-star winds becomes irrelevant to the process of
 BH X-ray binary formation, if the CE phase is initiated late in the
evolution of the massive primary, i.e., after core helium exhaustion. In
this case the helium star is exposed only through carbon burning and
later evolutionary phases. The total duration of these phases is so
short that the wind mass loss is insignificant and the helium-star mass
remains essentially constant~\cite{W95}. Current models of massive star
evolution~\cite{S92}, though, permit CE evolution at such late stages
only for primary masses lower than $\sim 25$\Msun\, and for an extremely
narrow range of orbital separations~\cite{KW98}. For more massive stars,
there is not enough radial expansion (in fact the radius decreases) to
counterbalance the orbital expansion due to wind mass loss from the
hydrogen-rich primary during core helium burning, and bring the primary
to Roche-loe overflow. Therefore, relying on CE evolution occurring only
at late stages cannot account for $\sim 10$\Msun\, BH and it is possible
only for a tiny fraction of primordial binaries leading to uncomfortably
low birth rates for BH X-ray binaries~\cite{PZ97}. All these problems
can be circumvented only if the radial evolution predicted by the
current models of mass-losing massive star evolution is incorrect both
qualitatively and quantitatively and stars more massive than $\sim
25$\Msun\, expand significantly after core helium exhaustion. Such a
modification of the massive-star models has been assumed by Wellstein \&
Langer~\cite{WL99} and no reduction of helium-star wind mass loss rates
is then required. Given the uncertainties in models of massive star
evolution such a modification cannot be excluded at present. In any
case, it becomes clear that the existence of X-ray transients with $\sim
10$\Msun\, BH requires that either the hydrogen-rich massive star models
or the strength of helium-star winds be modified.

\bigskip\noindent{\bf Acknowledgments.}
 I would like to thank N.\ Langer and S.\ Wellstein for useful
discussions. Support by the Smithsonian Astrophysical Observatory
through a Harvard-Smithsonian Center for Astrophysics Postdoctoral
Fellowship is also acknowledged.

\clearpage
\addcontentsline{toc}{section}{Index}
\flushbottom
\printindex

\end{document}